\documentclass{segabs}
\usepackage{mathtools}
\usepackage{amsfonts}
\usepackage{hyperref}
\DeclareMathOperator*{\argmin}{arg\,min}

\begin{document}


\onecolumn 

\begin{itemize}
\item[\textbf{Citation}]{A. Mustafa, M. Alfarraj, and G. AlRegib, “Spatiotemporal Modeling of Seismic Images for Acoustic Impedance Estimation,” Expanded Abstracts of the SEG Annual Meeting , Houston, TX, Oct. 11-16, 2020.}


\item[\textbf{Review}]{Date of presentation: 14 Oct 2020}




\item[\textbf{Copyrights}]{This is a preprint of a manuscript that has been accepted to be presented at the SEG Annual Meeting 2020. This preprint may strictly be used for research purposes only.}

\item[\textbf{Contact}]{\href{mailto:amustafa9@gatech.edu}{amustafa9@gatech.edu}  OR \href{mailto:alregib@gatech.edu}{alregib@gatech.edu}\\ \url{http://ghassanalregib.info/} \\ }
\end{itemize}


\thispagestyle{empty}
\newpage
\clearpage
\setcounter{page}{1}

\twocolumn
\title{Spatiotemporal Modeling of Seismic Images for Acoustic Impedance Estimation}

\renewcommand{\thefootnote}{\fnsymbol{footnote}} 

\author{Ahmad Mustafa$^{1}$\footnotemark[1], Motaz Alfarraj$^{2}$,
  and Ghassan AlRegib$^{1}$, Center for Energy and Geo Processing (CeGP), School of Electrical and Computer Engineering, Georgia Institute of Technology$^{1}$, \\King Fahd Univeristy of Petroleum and Minerals, Dhahran, Saudi Arabia.$^{2}$}

\footer{Example}
\lefthead{Mustafa \& AlRegib}
\righthead{Spatiotemporal Modeling for Seismic Inversion}

\maketitle

\begin{abstract}
Seismic inversion refers to the process of estimating reservoir rock properties from seismic reflection data. Conventional and machine learning-based inversion workflows usually work in a trace-by-trace fashion on seismic data, utilizing little to no information from the spatial structure of seismic images. We propose a deep learning-based seismic inversion workflow that models each seismic trace not only temporally but also spatially. This utilizes information-relatedness in seismic traces in depth and spatial directions to make efficient rock property estimations.  We empirically compare our proposed workflow with some other sequence modeling-based neural networks that model seismic data only temporally. Our results on the SEAM dataset demonstrate that, compared to the other architectures used in the study, the proposed workflow is able to achieve the best performance, with an average $r^{2}$ coefficient of 79.77\%. 
\end{abstract}

\section{Introduction}
\label{sec:intro}

Seismic inversion refers to the process of estimating reservoir rock properties from seismic reflection data. This allows the building of accurate and reliable subsurface models for oil and gas exploration. While the rock properties can be measured directly at the wells, they have to be estimated away from the well locations using seismic data. 

Classical seismic inversion is initiated with a smooth subsurface model. A synthetic seismic response is obtained from the initial model in a process called forward modelling. The resulting synthetic seismic response is compared to the actual seismic, and the error is used to update the model parameters. Many iterations of this process are performed until the synthetic seismic matches the actual seismic response to an acceptable degree of accuracy. This optimization procedure can be mathematically expressed as follows: 

\begin{equation}
    \hat{m} = \argmin_{m} \quad \mathcal{L}(f(m), d) + \lambda\mathcal{C}(m),
    \label{eq:1}
\end{equation}
where $f( m)$ represents the synthetic seismic generated by forward modelling on the model parameters, $\mathcal{L}(f(m), d)$ represents a distance measure between the synthetic seismic and the actual seismic $d$, $C(m)$ represents a regularization term imposed upon the problem to deal with the non-uniqueness of the solutions, $\lambda$ is the regularization weight, and $\hat{m}$ is the optimal solution found for the optimization problem. A comprehensive survey of the various seismic inversion methods is given in \cite{Veeken2004SeismicIM}. 

The use of machine learning and image processing algorithms to solve problems in seismic interpretation has become an active area of research. Machine learning has been used to solve problems in salt body delineation \citep{haibinSaltbodyDetection, AsjadSaltDetection, AmirSaltDetection} , fault detection \citep{haibinFaultDetection, HaibinFaultDetection2}, facies classification \citep{YazeedFaciesClassification, YazeedFaciesWeakClassification}, and seismic image retrieval and segmentation \citep{YazeedStructurelabelPrediction}. 

Machine learning algorithms have also been used to estimate physical properties of rocks \citep{ALANAZI201264, RothInversion, calderonParameterEstimation}. More recently, we saw Convolutional Neural Networks (CNNs) being used for rock property estimation from seismic data \citep{BiswasPhysicsGuidedCNN, DasCNNInversion}. Around the same time, \cite{motazRNN1} showed that by capturing the temporal relationships in seismic traces, Recurrent Neural Networks (RNNs) were able to efficiently estimate rock properties without actually requiring large amounts of training data, as is common with other non-sequence modelling based neural network architectures. Shortly afterwards, \citep{mustafaTCN} introduced another kind of sequence modelling neural network based on Temporal Convolutional Network (TCN) for estimation of acoustic impedance (AI) from seismic data. \citep{motazSemiSupervisedAcoustic, motazSemiSupervisedElastic} also showed that incorporating the forward model into the network architecture resulted in an implicit regularization of the network, thereby improving the quality of property estimations. 

A major drawback faced by classical and deep learning-based seismic inversion workflows is that each seismic trace is inverted independently of other traces. However, in a seismic image of the subsurface, neighbouring traces are highly correlated. A property estimation approach working on a trace-by-trace basis is not able to take this information into account. This can lead to lateral discontinuities in the inverted property volumes, especially in the presence of noise and large geological variations. However, naively extending a neural network architecture not specialized to handle spatial correlation to include neighboring seismic traces as features might not improve the network estimations. This is because the network does not pay regard to the ordering of the traces in the seismic image; it might even make matters worse since now the network has more parameters to learn. 

We propose a unique deep learning based seismic inversion workflow that models seismic data both \emph{temporally} and \emph{spatially}. Our network is derived from the CNN-based sequence modelling architecture described in \cite{Bai}, called a Temporal Convolutional Network (TCN). We extend and build upon this to introduce a two dimensional TCN-based architecture that is able to not only learn temporal relationships within each seismic trace, but also inject spatial context from neighboring traces into the network estimations. It does this by processing each data instance as a rectangular patch of seismic image centered at the well position, rather than just the single seismic trace at the well. By processing seismic data in this way, we are able to preserve and utilize the spatial structure of seismic images and still be able to model temporal relationships in seismic traces.  

The rest of the paper is structured as follows: we give a very brief overview of sequence modeling followed by a description of our Architecture. Next, we explain the set up of the problem. Lastly, we describe the dataset and our results, followed by a short conclusion.

\section{Sequence Modeling}
Let $\{x(0),...,x(T-1)\}$ and $\{y(0),...,y(T-1)\}$ be sequences respectively of the same length, where $T$ is the total number of time steps. A specific kind of sequence modeling is where a point in the latter sequence at time $T$, $y(T)$ depends only on the samples of the former sequence $x(t)$ for $t \leq T$. This mapping described by Equation \ref{eq:seq} can be represented by a neural network $\mathcal{F}$ parameterized by $\Theta$ (i.e., $\mathcal{F}_{\Theta}$).
\begin{equation}
    y(t) = \mathcal{F}_{\Theta}\left(x(0),\dots,x(t)\right) \forall t\in [0,T-1].
    \label{eq:seq}
\end{equation}

Historically, Recurrent Neural Networks (RNNs) have featured extensively in various sequence modeling tasks. Their popularity comes from the fact that they can maintain a vector of hidden states through successive points in time, allowing them to use history of past inputs to make predictions at future times. Over time, several variants of the original RNN architecture, like LSTMs and GRUs were proposed that were easier to train and modelled long term dependencies better. \citep{lipton2015critical} give a detailed review of RNNs for sequence learning tasks. Convolutional Neural Networks (CNNs) have also been used extensively for sequence modeling tasks like document classification \citep{johnson}, machine translation \citep{Kalchbrenner}, audio synthesis \citep{Oord}, and language modeling \citep{Dauphin}.

\section{Network Architecture}
\begin{figure*}
    \centering
    \includegraphics[width=2\columnwidth]{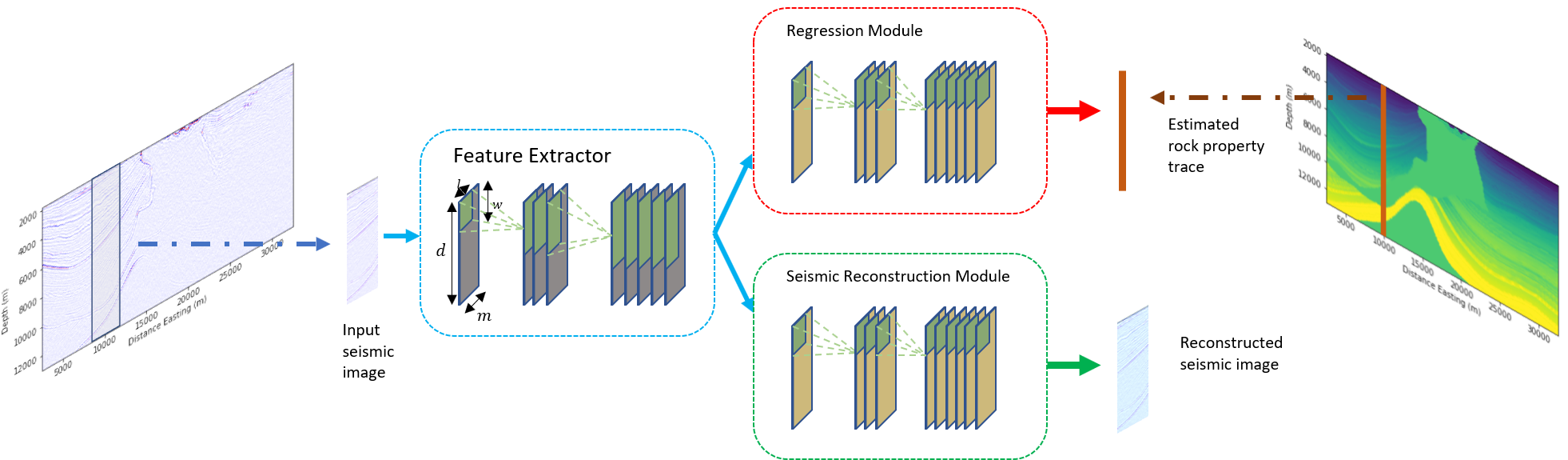}
    \caption{The proposed network architecture. A seismic image of dimensions $d\times m$ is processed by the feature extractor block that consists of multiple 2-D temporal blocks. The 2-D kernels stay fixed in length $l$ but increase exponentially in width $w$. The features extracted by this block are output simultaneously to two different shallow 2-D CNNs to output estimated property trace and reconstructed input seismic image respectively.  }
    \label{fig:network}
\end{figure*}

The network architecture is shown in Figure \ref{fig:network}. As described in \cite{mustafaTCN}, the main body of the network, the feature extractor is composed of multiple temporal blocks, except that these are based on 2-D rather than 1-D convolutions. 
The input to the network is a rectangular patch of seismic data centered at the well position. The feature extractor block in the network processes this seismic image using 2-D kernels in each layer. The kernel width stays constant while the height increases due to dilation. This lets seismic traces be modelled temporally to efficiently learn the well log property at that particular well position while also injecting spatial context into network estimations. The number of channels is increased after each layer, starting at 1 at the input layer to 120 at the last layer of the feature extractor. Increasing the number of channels allows the network to learn more features. The output of the feature extractor block is fed simultaneously into a 'Regression module' and a 'Seismic Reconstruction Module'. Both of these consist of shallow  2-D convolutional networks made up of 3 layers each. The regression module  processes the activations  to produce an estimate of the well-log property. The 'Seismic Reconstruction Module' uses the activations generated by the Feature Extractor to reconstruct the input seismic image. This is performed as a form of regularization to produce more stable network estimations.

\section{Methodology}
Consider $\mathcal{D} = \{\mathcal{X}, \mathcal{Y}\}$ to represent the dataset. $\mathcal{X} = \{x^{1}, ..., x^{N}| x^{i} \in \mathbb{R}^{d \times m}\}$ represents the collection of $N$ seismic images in a dataset, where each $x^{i}$ is a $d\times m$ dimensional image. $d$ refers to the depth of the image while $m$ is the width. $\mathcal{Y} = \{y^{1}, ..., y^{N}|y^{i}\in\mathbb{R}^{d}\}$ refers to collection of well log properties corresponding to each $x^{i} \in \mathcal{X}$, where each $y^{i}$ is a $d$ dimensional rock property trace.
A batch of seismic images from the dataset is processed to get the estimated well properties as well as the reconstructed seismic image as shown below:

\begin{equation}
    \hat{y}^{i}, \hat{x}^{i} = \mathcal{F}_{\Theta}(x^{i}),
\label{eq:2}
\end{equation}
where $\hat{y}^{i}$ and $\hat{x}^{i}$ are, respectively, the estimated well log property and the reconstructed seismic image at the output of the network $\mathcal{F}$ characterized by its weights $\Theta$ from the seismic input $x^{i}$. The training process can then be summarized as:

\begin{equation}
    \hat{\theta} = \argmin_{\Theta} 
    \quad  \frac{1}{|\mathcal{D}|}\sum_{(x^{i},y^{i})\in\mathcal{D}} \alpha\times\\|\hat{y}^{i} - y^{i}\|^{2}_{2} + 
    \beta\times \|\hat{x}^{i} - x^{i}\|^{2}_{2}. 
\label{eq:3}    
\end{equation}

For each iteration, the network compares the estimated well properties to the ground truth properties, reconstructed seismic images with the input images, and sums these losses over all the training examples in the dataset. We chose $\alpha = 1$ and $\beta = 0.5$. The seismic reconstruction loss $ \|\hat{x} - x\|^{2}_{2}$ is given a weight of 0.5 so as to give more importance to the well-log estimation loss $\|\hat{y} - y\|^{2}_{2}$. The losses are then backpropagated into the network  to decrease the loss at the next iteration by changing the weights accordingly.

\section{Results and Discussion}

\begin{figure*}
    \centering
    \includegraphics[width=\textwidth]{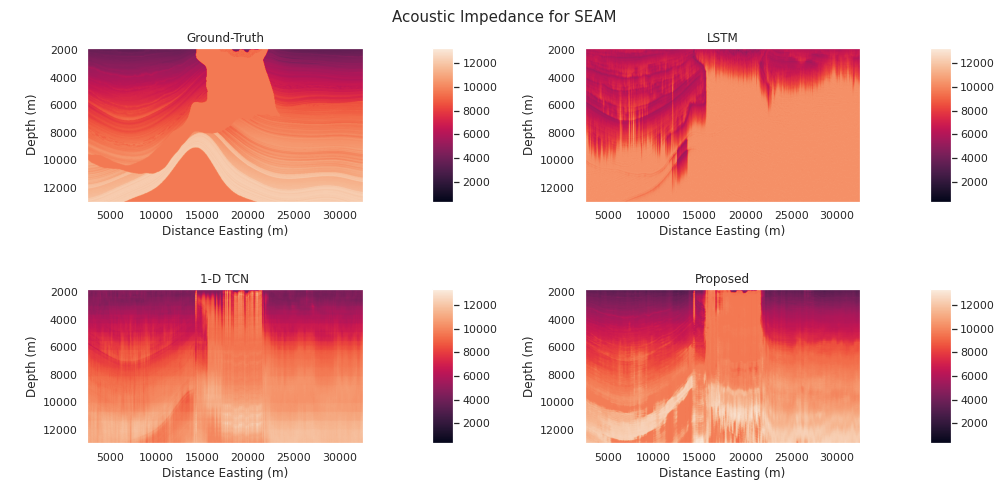}
    \caption{Ground truth AI sections (top left),  estimated by the 1-layer LSTM (top right), 1-D TCN (bottom left), and the proposed 2-D TCN based architecture (bottom right). Notice that the proposed approach leads better delineation of the various subsurface structures, especially the salt in the bottom left of the section. The LSTM performed much poorly than the other architectures, with unstable training loss and converging to a higher final loss value.}
    \label{fig:sections}
\end{figure*}

\begin{figure}
    \centering
    \subfigure[x=5000m]{\includegraphics[width=0.3\columnwidth]{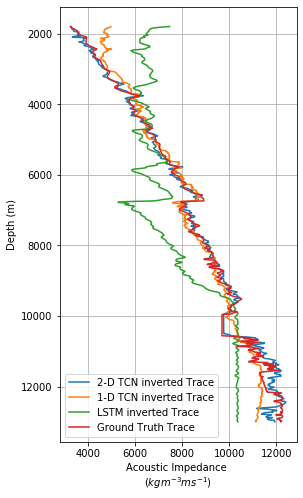}}
    \subfigure[x=18000m]{\includegraphics[width=0.3\columnwidth]{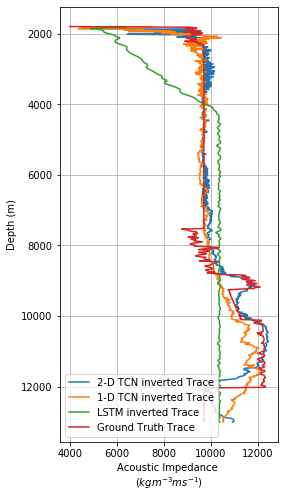}} \\
    \subfigure[x=30000m]{\includegraphics[width=0.3\columnwidth]{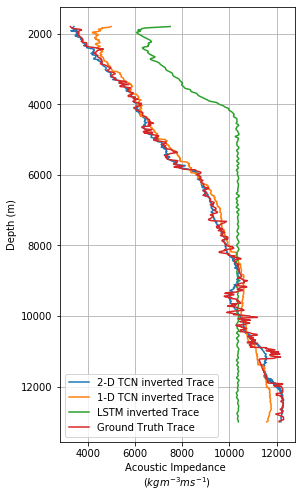}}
    \caption{Pseudologs at a) x=5000m, b) x=18000m, and c) x=30000m along east. It can be seen that the AI trace estimated with the proposed workflow (blue) corresponds better to the ground-truth (green) than the one obtained simply by a 1-D TCN (orange).}
    \label{fig:traces}
\end{figure}

We validate our architecture on the opensource SEAM dataset. It comes with a 3-D seismic survey and density and p-velocity models along one cross-section of the survey. We obtain the ground truth p-impedance model by multiplying the density by the p-velocity. This 2-dimensional model cross-section can also be considered a collection of Acoustic Impedance pseudologs for each of whom a corresponding seismic trace is also available in the survey. We uniformly sample the logs and pick out one approximately every 2.1km. This provides a total of 14 wells over a distance of 30km, which is less than 3\% of the total available logs. For each well, we also sample a seismic image seven traces wide centered at the well position. The training is carried out using the popular Python-based deep learning framework \emph{PyTorch}. We train our proposed model for 1000 epochs with the Adam optimization algorithm. We use a batch size of 14, initial learning rate of 0.001, and a weight decay of 0.0001 to counter overfitting. Since our input i.e. seismic data, are images, we also perform data augmentation by random horizontal flipping of images.

To compare our proposed workflow with sequence modelling architectures without spatial awareness, we also train separately, using the same settings, a 1-D TCN and a one layer LSTM-based architecture. The three trained models are then tested on all seismic traces in the dataset to give estimated acoustic impedance sections, which are shown in Figure \ref{fig:sections}. As can be seen in the figure, the 2-D TCN-based architecture performs better at estimating the acoustic impedance of the seismic section compared to the other two architectures. It delineates the salt region in the bottom left of the section much more clearly compared to the 1-D TCN and LSTM. The LSTM converged to a significantly higher loss value compared to the first two networks. It completely misses the low frequency trend even though in some places, it retains the high frequency characteristics of the seismic data. This can also be seen in the trace plots for the different networks at 3 different positions in Figure \ref{fig:traces}. The blue trace corresponding to the estimated AI by 2-D TCN corresponds much better to the ground truth (red) compared to the other two. While the 1-D TCN is able to match the ground truth for the most part reasonably well, it is thrown off by the sudden increase in AI due to the salt around a depth of 10000m. The blue trace can be seen to do much better around these depths. Finally, to measure the overall regression performance in terms of quantitative metrics, we also compute and present in the table below the average $r^{2}$ coefficient and Pearson's Correlation Coefficient (PCC) over all the AI pseudologs in the seismic section. 
   
\begin{table}[ht]
\centering
 \begin{tabular}{c|c|c|c} 
 \hline
\textbf{ Metric} & \textbf{Proposed} & \textbf{1-D TCN} & \textbf{LSTM}  \\
 \hline
 \hline
 PCC & \textbf{0.9259} & 0.9088 & 0.8526 \\ 
 \hline
 $r^{2}$ & \textbf{0.7977} & 0.6709 & 0.4738\\
 \hline
\end{tabular}
\caption{Performance metrics for all three architectures used in the study evaluated over the whole seismic section.}
\label{tab:results}
\end{table}

\section{Conclusion}
Conventional sequence modelling based neural networks for seismic inversion suffer the limitation of only modeling seismic data temporally. This ignores the information embedded in the spatial structure of seismic images. We describe a sequence modeling approach that is able to model spatiotemporal data like seismic images to better estimate outputs. We validate our approach by performic estimation of Acoustic Impedance on the SEAM dataset, where our algorithm performs better than the 1-D sequence modelling networks used in the study. 
\bibliographystyle{seg}
\bibliography{example}

\end{document}